\documentclass[conference]{IEEEtran}
\usepackage[tight,footnotesize]{subfigure}
\usepackage{url}
\usepackage{listings}
\usepackage{xcolor}
\usepackage{hyperref}
\usepackage{booktabs}

\usepackage{balance}

\colorlet{punct}{red!60!black}
\definecolor{background}{HTML}{EEEEEE}
\definecolor{delim}{RGB}{20,105,176}
\colorlet{numb}{magenta!60!black}

\makeatletter
\def\lst@makecaption{%
  \def\@captype{table}%
  \@makecaption
}
\makeatother

\usepackage{tikz}
\usepackage{xcolor}
\usepackage{color}
\usepackage{pgfplots}

\usetikzlibrary{
	arrows.meta,
	calc,
	colorbrewer,
	external,
	fit,
	matrix,
	positioning,
	shapes,
	shapes.misc,
	backgrounds
}

\begin{document}

%%% title and authors
%%\title{Report on LoRaWAN field test and deployments in MONICA using RIOT-OS and TTN (to-be-changed)}
%\title{On the Suitability of a Public LoRaWAN:\\ Insights from Field Tests in the MONICA Project}
%\title{MONICA on LoRaWAN: Insights from Field Deployments of Smart City IoT Applications}
\title{LoRa in the Field:  Insights from Networking the Smart City Hamburg with RIOT}
%\author{
%	\IEEEauthorblockN{
%		Sebastian Meiling\IEEEauthorrefmark{1},
%		Dorothea Purnomo\IEEEauthorrefmark{3},
%		Julia-Ann Shiraishi\IEEEauthorrefmark{2},
%		Michael Fischer\IEEEauthorrefmark{3}, and
%		Thomas C. Schmidt\IEEEauthorrefmark{1}
%	}
%	\vspace{.5ex}
%	\IEEEauthorblockA{
%		\IEEEauthorrefmark{1}\{sebastian.meiling, t.schmidt\}@haw-hamburg.de, 
%		\IEEEauthorrefmark{2}julie-ann.shiraishi@sk.hamburg.de,\\
%		\IEEEauthorrefmark{3}\{dorothea.purnomo, michael.fischer1\}@gv.hamburg.de
%	}
%	\vspace{.5ex}
%	\IEEEauthorblockA{
%		\IEEEauthorrefmark{1}iNET RG, Hamburg University of Applied Sciences, Germany}
%	\IEEEauthorblockA{
%		\IEEEauthorrefmark{2}Senate Chancellery, Free and Hanseatic City Hamburg, Germany}
%	\IEEEauthorblockA{
%		\IEEEauthorrefmark{3}Agency for Geoinformation and Surveying, Free and Hanseatic City Hamburg, Germany}
%}
\author{
	\IEEEauthorblockN{
		Sebastian Meiling,
		Thomas C. Schmidt
	}
	\vspace{.5ex}
	\IEEEauthorblockA{
		\{sebastian.meiling, t.schmidt\}@haw-hamburg.de
	}
	\vspace{.5ex}
	\IEEEauthorblockA{
		iNET RG, Hamburg University of Applied Sciences, Germany
	}
}
%% copyright statement
\IEEEoverridecommandlockouts
%\IEEEpubid{978–1–5386–1478–5/18/\$31.00~\copyright~2018 IEEE}
\IEEEpubid{}
%% make title
\maketitle

%%% sections
\begin{abstract}
Inter-connected sensors and actuators have scaled down to small embedded devices such as wearables, and at the same time meet a massive deployment at the Internet edge---the Internet of Things (IoT).
Many of these IoT devices run on low-power batteries and are forced to operate on very constrained resources, namely slow CPUs, tiny memories, and low-power radios.
 Establishing a network infrastructure that is energy efficient, wireless, and still covers a wide area is a larger challenge in this regime.
 LoRa is a low complexity long range radio technology, which tries to meet these challenges.
 With LoRaWAN a network model for widespread deployment has been established, which enjoys open public LoRaWAN dissemination such as with the infrastructure of TheThingsNetwork.

In this paper, we report about our experiences with developing and deploying LoRa-based smart city applications as part of the MONICA project in Hamburg. Our contributions are twofold.
First, we describe the design and implementation of end-to-end IoT applications based on the friendly IoT operating system RIOT.
Second, we report on measurements and evaluations of our large field trials during several public events in the city of Hamburg.
Our results show that LoRaWAN provides a suitable communication layer for a variety of Smart City use-cases and IoT applications, but also identifies its limitations and weaknesses.
\end{abstract}

\section{Introduction}
\label{sec:intro}

The Internet of Things (IoT) has introduced the technologies to connect an unprecedented number of (smart) electronic devices with the global Internet.
The IoT thus opened the opportunity to distribute smart sensors and actuators in the field, or to augment existing controllers with Internet connectivity.
A large, emerging deployment area are smart cities, in which a distributed system of intelligent appliances shall increase comfort and safety for  citizens.
Hamburg has taken up early the opportunity toward a smarter city life  and strategically explores how to make urban data beneficially utile to the public.
The EU project MONICA, which explores the large-scale deployment of smart wearables in public spaces, is part of this initiative.

Wearable IoT appliances are typically small, embedded devices with constrained resources, i.e., low energy, battery powered, less memory and slower CPUs compared to standard PCs.
Due to this nature, any communication technology suitable for the IoT should be energy efficient, secure, wireless, and cheap.
Achieving these key requirements typically comes at the cost of availability, bandwidth, latency, and robustness.   
Connectivity of IoT devices can be achieved by a multitude of standards and technologies.

The most common communication technologies can be roughly categorised in terms of its range: (a) short-range such as  \emph{Bluetooth Low Energy} (BLE), \emph{Near Field Communication} (NFC), and RFID; (b) mid-range, e.g. IEEE 802.15.4 and Zigbee; and long range, e.g., \emph{Narrow-Band IoT} (NB-IoT), \emph{SigFox}, and \emph{LoRa/LoRaWAN}~\cite{vlnkm-ccasl-17}.
In general, only the latter are suitable to build a city wide network infrastructure achieving good coverage and (ideally) low costs.
As of today, though, only few evaluations of real world IoT applications and deployments in Smart City environments are available that use low-power, long-range radio technologies.

Our key contribution in this paper are:
First, we describe the implementation and deployment of 2 distinct end-to-end IoT applications, namely GPS Trackers and environmental sensors.
Second, we present the evaluation results of measurements gathered during the field deployments at pilot events of the MONICA project.
Our IoT devices use the public LoRaWAN infrastructure of TheThingsNetwork\footnote{https://www.thethingsnetwork.org} for communication and run a firmware based on RIOT-OS~\cite{bghkl-rosos-18}, an Open Source operating system for the IoT.
The application backend utilises the IoT platform developed in the MONICA project.

The remainder of this paper is structured as follows. We provide background information and an overview on related work in \autoref{sec:background}.
In \autoref{sec:monica}, we introduce use cases and events of the MONICA project, which were used for demonstration in the Hamburg pilot.
Section \ref{sec:deploy} describes the setup and deployment of the pilot demonstrations.
We present and discuss our evaluation results in \autoref{sec:eval}.
Finally, we conclude and give an outlook on future work in \autoref{sec:outlook}.  

\section{Background and Related Work}
\label{sec:background}
 
In our previous work~\cite{mpsfs-mhtli-18} we investigated end-to-end IoT communication based on IoT standard protocols and technologies such as LPWAN with IEEE 802.15.4, 6LoWPAN/IPv6, UDP, and CoAP.
In this paper we continue and extend this work with a different deployment setup based on LoRaWAN for communication. 
\pagebreak

% LoRa(WAN) related work
The latest specification of LoRaWAN is version 1.1~\cite{lorawan-spec-11}. 
LoRaWAN is based on the LoRa radio standard as its physical layer~\cite{la-toll-15}.
LoRa utilises unlicensed frequency bands~\cite{cvzz-lcubr-16} and is based on a chirp spread spectrum (CSS) modulation~\cite{IEEE-802.15.4-16}.

%% theory, no deployment
The work of Silva et al.~\cite{srasa-llpwp-17} gives an overview on the challenges and opportunities of LoRaWAN.
They show that LoRaWAN can provide a suitable communication infrastructure for low-power, embedded IoT devices with high energy efficiency and long range coverage.
In \cite{avtms-ull-17} Adelantado et al. analyse limitations of LoRa(WAN) with respect to potential use-cases.
They argue the most restricting factor for communication is the required duty cycle for radio access that also affects throughput and capacity.
The work of Slabicki et al.~\cite{spd-aclnd-18} is twofold: they contribute a LoRa simulation framework for OMNeT++, and present an adaptive data rate for LoRa to optimise bandwidth utilisation.

% with deployment
Liando et al.~\cite{lgtl-kufle-19} investigate the energy efficiency and communication xperformance of LoRa.
They found that LoRa can achieve radio coverage beyond 10km given line-of-sight, but suffers from degradation in dense city environments where buildings obscure communication between IoT devices and LoRa gateways.
An evaluation of the indoor and outdoor performance of LoRaWAN was done by Wixted et al.~\cite{wklta-ellws-16} in the City of Glasgow. 
They show that LoRaWAN can provide reliable communication layer for sensor applications. 
In their paper~\cite{ypmp-il5tn-17} Yasmin et al. present the integration of LoRaWAN with a 5G test network.
They argue that upcoming 5G networks with high bandwidth and capacity can serve as a backbone network to connect LoRaWAN gateways in the field with the global Internet.
Blenn et al.~\cite{bk-lwmtn-17} present an early measurement study of the TheThingsNetwork LoRaWAN infrastructure.
They analyse the effect of payload size and signal quality on the network performance.

\section{MONICA: Use Cases and Pilot Events}
\label{sec:monica}

The EU project MONICA\footnote{https://www.monica-project.eu} aims at solving typical problems of large inner-city public events.
Key challenges on that matter are crowd safety and security as well as sound and noise pollution.
The main objective of MONICA is to utilise multiple existing and new IoT technologies and show-case solutions in large scale pilot demonstrations.
With that the project is also part of and contributes to the IoT European Large-Scale Pilots Programme (LSP)\footnote{www.european-iot-pilots.eu}.
A core component of the MONICA project is its IoT platform and ecosystem that enables a variety of IoT applications to address issues associated with public events, such as funfairs or festivals. 
More information and details on the MONICA IoT platform and its components can be found in~\cite{mpsfs-mhtli-18}.

The MONICA project consortium consists of 29 partners from 9 EU countries, with pilot demonstrations at several events in 6 European cities.
The MONICA project defines a variety of use case groups to address (and solve) specific aspects of the aforementioned challenges for large public events.
Each pilot city selects a subset of these use cases which are most suitable to the needs and requirements of their respective pilot events.
In the following we describe two use cases for the Hamburg pilot demonstrations and give an overview on the pilot events.
The implementation and deployment setup of these use-cases is described afterwards in \autoref{sec:deploy}. 

\begin{figure}[t]
    \centering
    \includegraphics[width=\columnwidth]{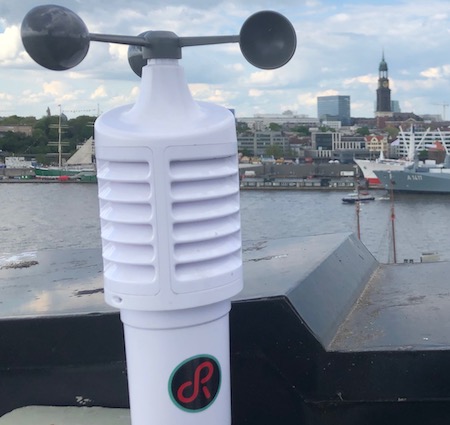}
    \caption{Photo from a windspeed sensor deployed at the Port Anniversary pilot demonstration.}
    \label{fig:deploy_sensor}
\end{figure}

\subsection{Use Cases}
\label{ssec:use_cases}

The first use case is \emph{Locate Staff} that covers the MONICA solutions for locating and monitoring  event staff members.
The main goal here is the management of (human) resources by providing an overview of the location of actors as well as associated meta data and time-related (historic) information.
Specific tasks are: (a) monitor the location, state and availability of staff members; (b) locate staff members who are closest to security incidents.
This solution is easily integrated with other MONICA use cases such as crowd management, incident handling and more.

Second, is the \emph{Security and Health Incidents} use case group that aims to support the detection, reporting and handling of health, security and safety incidents that may occur during public events.
The general goal is that the MONICA platform is able to automatically detect incidents or potential threats through different IoT technologies such as smart cameras, microphones, and environmental sensors.
The data is processed by the MONICA platform, i.e., its decision support system (DSS), and the outcome is visualised through dedicated user applications such as the COP (common operational picture).
Overall this solution aids the event staff in detecting and handling incidents.

\subsection{Pilot Events}
\label{ssec:events}

The Free and Hanseatic City Hamburg is one of six pilot cities in the MONICA project to run large-scale IoT deployment demonstrations.
Hamburg participates with two of its most prominent annual public events, namely the Hamburger DOM funfair and the Port Anniversary festival.

The \emph{Hamburger DOM} is Northern Germany’s biggest funfair with 7--10 million annual visitors during the 91 DOM days.
The funfair takes place in the premises of the Heiligengeistfeld with a total of around 250 rides, stalls, and attractions.
The port of Hamburg is the most important port in Germany, and a leading cargo handling centre in the world. 
Each year, more than one million visitors from Germany and abroad come to the \emph{Hamburg Port Anniversary} to join the atmosphere created by ships from all parts of the worlds. 
The festival area extends six kilometres along the waterfront and includes displays on both land and water.  
A major attraction are several ship parades during the event.

\begin{figure}[t]
    \centering
    \includegraphics[width=\columnwidth]{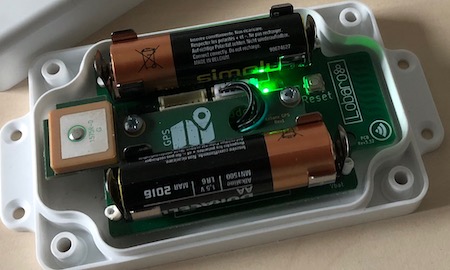}
    \caption{Photo from a look into to hardware of the GPS tracker devices used at the Hamburger DOM funfair.}
    \label{fig:deploy_tracker}
\end{figure}

\section{Implementation and Deployment}
\label{sec:deploy}

The two MONICA use-cases (locate staff and security incidents) are implemented by separate end-to-end IoT applications ---i.e., GPS tracking and environmental sensors--- using open IoT technologies and protocols.
This includes (a) the public LoRaWAN infrastructure provided by \emph{TheThingsNetwork}, i.e., their LoRaWAN gateways, network and application servers; (b) the Open Source IoT operating system RIOT-OS~\cite{bghkl-rosos-18}; and (c) the MONICA IoT platform to store, process, and access the gathered data and derived information.
The key hardware and software components of both applications are described here.

Note that we have chosen to use LoRa/LoRaWAN for our deployments over other technologies, i.e., Sigfox and NB-IoT, for several reasons.
Most importantly: there is an active LoRaWAN community and growing infrastructure of TTN in Hamburg and the services are provided openly and free of charge.

\subsection{IoT Device Implementation}
For the GPS tracker application we use an embedded, wearable hardware platform based on the low-power STM32 L151 MCU.
The platform is equipped with GPS module for localisation and a LoRa modem for communication.
Power is provided by two standard AA batteries (see \autoref{fig:deploy_tracker}).
On the software side we developed our own custom firmware based on RIOT-OS to have full control over message format and data interval.
The \emph{locate staff} use case required to have precise location information in regular time intervals, for the pilot demonstrations we aimed for a message interval of 30\,s.

For the environmental sensors we also used an embedded hardware platform based on the STM LoRa development kit B-L072Z-LRWAN1\footnote{https://www.st.com/en/evaluation-tools/b-l072z-lrwan1.html} that has a LoRa modem integrated.
Further, we equipped it with an off-the-shelf sensor that provides measurements of air humidity, temperature, and wind speed (see \autoref{fig:deploy_sensor}).
As before we developed our own custom firmware based on RIOT-OS to periodically send sensor values via LoRaWAN to the MONICA backend.
For the \emph{Security Incidents} use case we aimed for a message interval of 60\,s.

The RIOT-OS based firmwares for the GPS trackers\footnote{https://github.com/MONICA-Project/lorawan-tracker} and the environmental sensors\footnote{https://github.com/MONICA-Project/lorawan-sensors} are published as Open Source on Github.

\subsection{LoRa Configuration}
For both applications it was necessary to match data interval requirement of the use-cases with the restriction imposed by LoRa.
Namely the duty cycle that limits the amount of data a device is allowed to send within a given time period.
To derive an optimal configuration we had two options for adjustment: the data size, i.e., message payload; and the spreading factor (SF).
The latter is a LoRa parameter that has direct influence on throughput and radio coverage.
A high spreading factor has less throughput but better coverage, and vice-versa.
Based on payload size and spreading factor we calculated the minimal allowed data interval as specified in~\cite{semtech-an1200-13}

\begin{table}[t]
	\centering
	\caption{Overview on the configuration parameters for the LoRaWAN communication for the GPS tracker and environmental sensor IoT applications.}
	\label{tab:deploy}
	\begin{tabular}{lrrrrr}
	\toprule
	    & {\bf Payload} & {\bf Spreading} & {\bf Duty } & {\bf minimal}  & {\bf target}\\
	    & {\bf Size}    & {\bf Factor}    & {\bf Cycle} & {\bf Interval} & {\bf Interval} \\
	\midrule
	GPS Trackers & 11\,Bytes & SF9        & 1\,\%       & 20.58\,s       & 30\,s\\
	Env. Sensors & 8\,Bytes  & SF10       & 1\,\%       & 37.07\,s       & 60\,s\\
	\bottomrule
	\end{tabular}
\end{table}

\begin{figure}[t]
	\centering
	\includegraphics[width=\columnwidth]{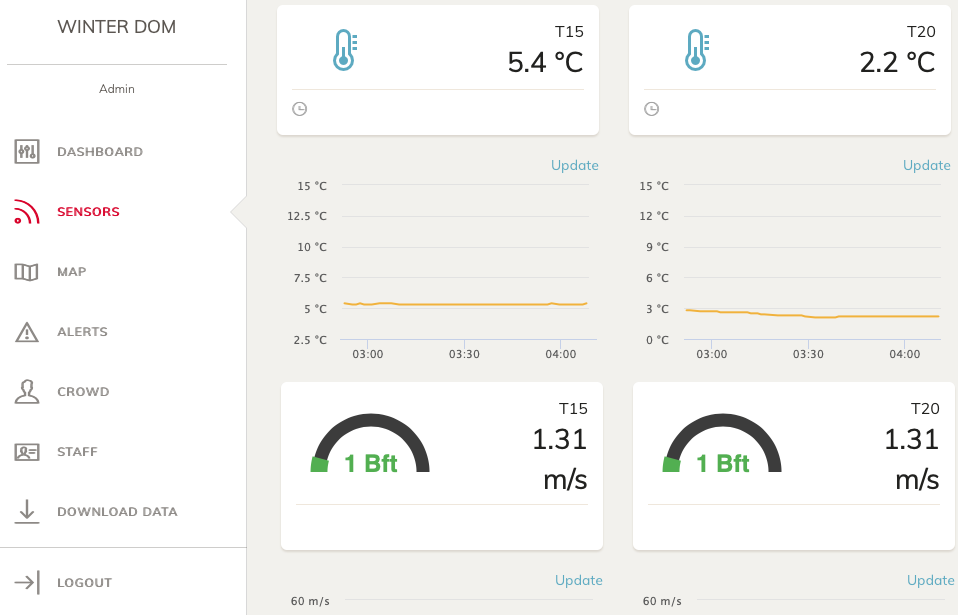}
	\caption{Screenshot of the MONICA COP showing sensor readings.}
	\label{fig:cop}
\end{figure}

\autoref{tab:deploy} gives on overview on the final parameter and settings for both applications.
Note, as the trackers required a shorter message interval we used SF9, compared to SF10 for the environmental sensors.

\pagebreak

\subsection{End-to-End Data Flow}
The end-to-end data flow of both application is very similar.
The tracker and sensor devices send data messages according to the defined timing interval via LoRaWAN.
Each TTN LoRaWAN gateway that receives an encrypted LoRa message forwards its associated LoRaWAN network server.
Note, gateways are simple forwarding entities, they do not access or decrypt any data message.
The network server decrypts the message header and forwards the message to the corresponding TTN application server.
Note, the network server cannot access or decrypt the message payload.
Afterwards the data (message payload) is send via HTTPS to a proxy of the MONICA platform that handles further processing and storage.
Finally the processed information is visualised in dedicated MONICA user applications, e.g., the common operational picture (COP), see \ref{fig:cop}.
Note, during each forwarding step the data is secured and only send encrypted.
The LoRaWAN part is secured by network and applications keys (see \cite{lorawan-spec-11}), the MONICA backend utilises \emph{Keycloak}\footnote{https://www.keycloak.org/} to generate keys for authentication and authorisation.
\section{Evaluation}
\label{sec:eval}

In our evaluation we focus on the influence of the LoRaWAN infrastructure on end-to-end metrics such as latency and robustness from an IoT application perspective.
Specifically, we analyse how the number of LoRaWAN gateways within transmission range of a device and their distance influences our metrics.
The evaluation is based on our measurements from the Hamburg pilot events, i.e., the Port Anniversary and the Hamburger DOM, and the aforementioned IoT applications, i.e., GPS trackers and environmental sensors.
Measurement data was gathered during 4 pilot demonstrations in 2019, i.e., 1x Port Anniversary in May, and 3x DOM funfair in March, July, and November.
In total, we analysed more than 295.000 LoRaWAN messages received during all pilot events.

\begin{figure}[t]
	\includegraphics[width=\columnwidth]{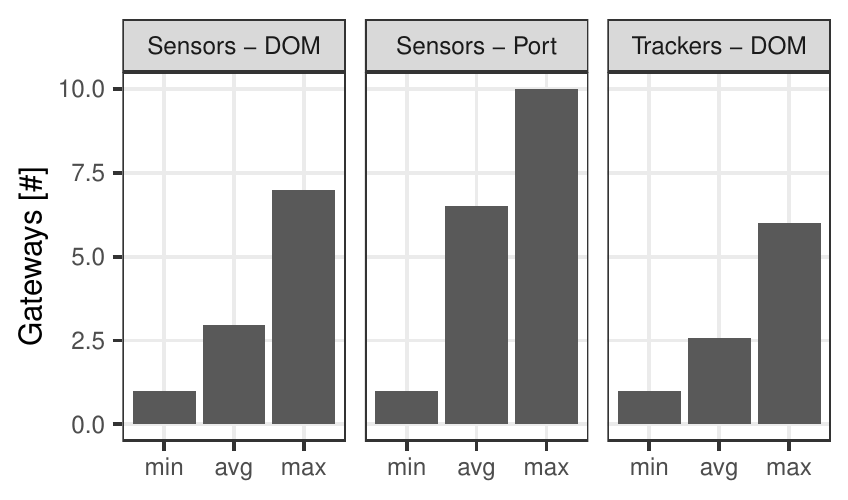}
	\caption{Comparing the minimal, average and maximum number of LoRaWAN gateways reached by all IoT devices during a specific MONICA pilot event.}
	\label{fig:gtws}
\end{figure}

The graph in \autoref{fig:gtws} gives an overview on the number of LoRaWAN gateways reachable by all IoT device during distinct events.
The sensors and trackers at the DOM reach fewer gateways on average compared to the sensors deployed at the port.
This is due to the denser urbanisation around the DOM event site, i.e., more and higher buildings that can block radio communication between devices and gateways.
We also see that the trackers have even lower values which is expected as they were carried by staff members at roughly 1\,m above ground, while the sensors were installed on poles in 5 - 10\,m height.

First, we look into the measurement results for the GPS tracker.
In the following three graphs we analyse the influence of the relative gateway position on the overall reliability of LoRaWAN data transmission.
Therefore we compare two deployments of the GPS trackers at the Hamburg DOM event, i.e., one with a gateway directly deployed at the event site; and second deployment without that gateway.

\begin{figure}[t]
	\includegraphics[width=\columnwidth]{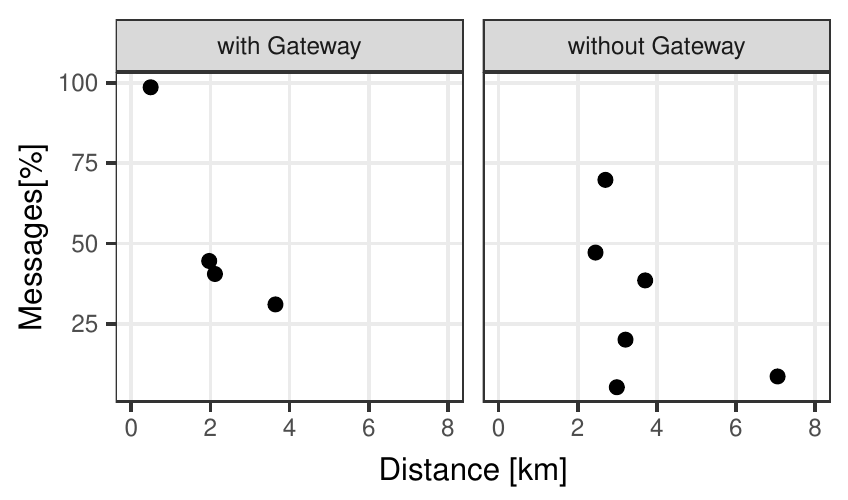}
	\caption{Comparison of gateway-to-device distance and amount of messages received by a certain gateway for the GPS trackers with and without gateway at the deployment site of the MONICA pilot event.}
	\label{fig:tracker_dist}
\end{figure}

\begin{figure}[t]
	\includegraphics[width=\columnwidth]{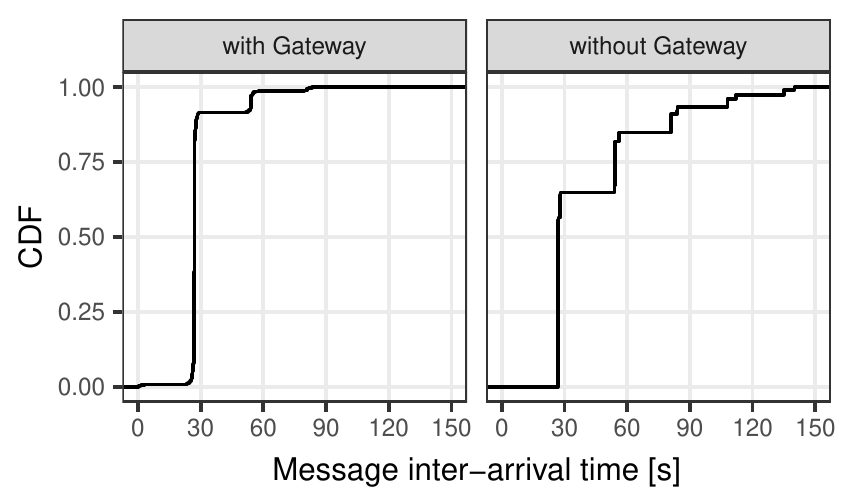}
	\caption{Comparison of the message inter-arrival times for the GPS trackers with a gateway (left) and without gateway (right) at the deployment site of the MONICA pilot event.}
	\label{fig:tracker_diff}
\end{figure}

\begin{figure}[t]
	\includegraphics[width=\columnwidth]{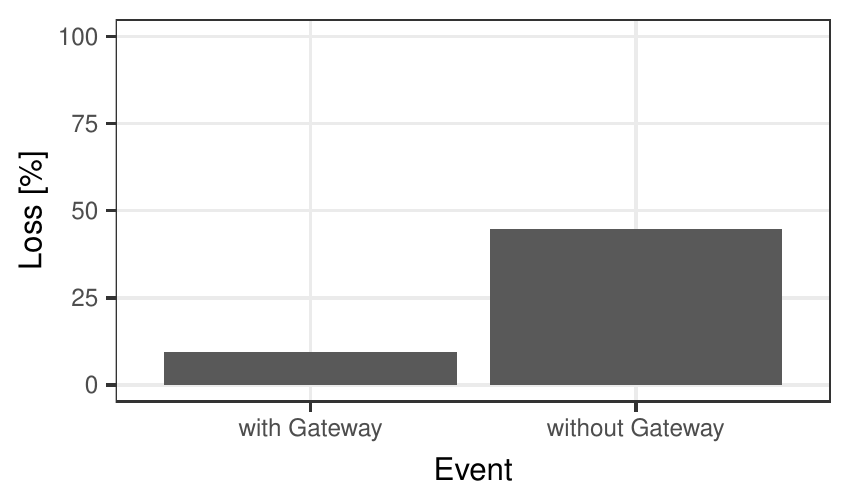}
	\caption{Comparison of the message loss for the GPS trackers with a gateway (left) and without gateway (right) at the deployment site of the MONICA pilot event.}
	\label{fig:tracker_loss}
\end{figure}

We compare the gateway-to-device distance and the relative amount of messages received by a certain gateway in \autoref{fig:tracker_dist}.
In the deployment with a gateway at the event site, we see that one gateway is less than 1\,km away and receives nearly 100\,\% of all messages, while the other gateways farther away only receive around 45\,\% of all messages at best.
For the second deployment without a gateway at the event site, we see a drop in performance.
The closest gateway is more than 2\,km away and receives $\approx$ 70\,\% of all messages.
This performance degradation is also reflected by the message inter-arrival times as shown in \autoref{fig:tracker_diff} and the summarised packet loss as shown in \autoref{fig:tracker_loss}.
Specifically, \autoref{fig:tracker_diff} shows that with a gateway around 90\,\% of all messages were received in the 1st interval (matching the targeted duty cycle of roughly 30\,s); and already 99\,\% were received within 60\,s.
Without a gateway this drops to 65\,\% for the 1st interval; and generally longer inter-arrival times, i.e., reaching 99\,\% reception rate after 120\,s.
The packet loss is $\approx$ 10\,\% for the deployment with a LoRaWAN gateway at the event and 45\,\% without, see \autoref{fig:tracker_loss}.

\begin{figure}[t]
	\includegraphics[width=\columnwidth]{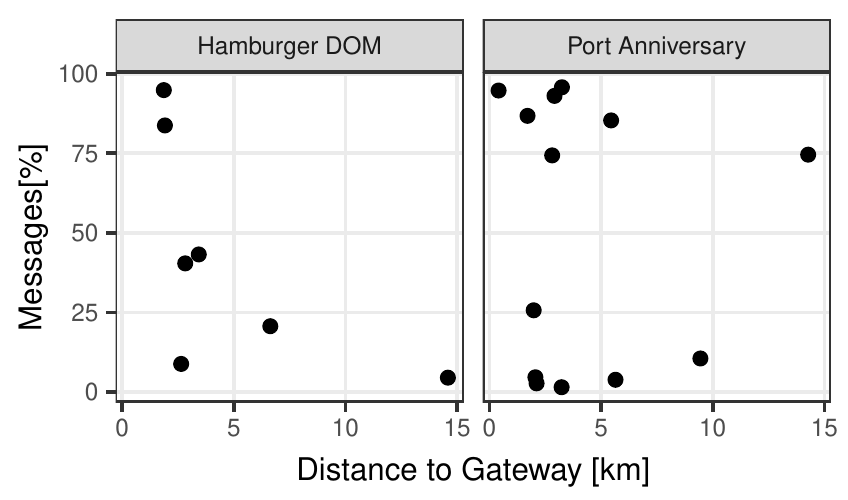}
	\caption{Comparison of gateway-to-device distance and amount of messages received by a certain gateway for environmental sensors at the Hamburger DOM and Port Anniversary MONICA pilot event.}
	\label{fig:sensor_dist}
\end{figure}

\begin{figure}[t]
	\includegraphics[width=\columnwidth]{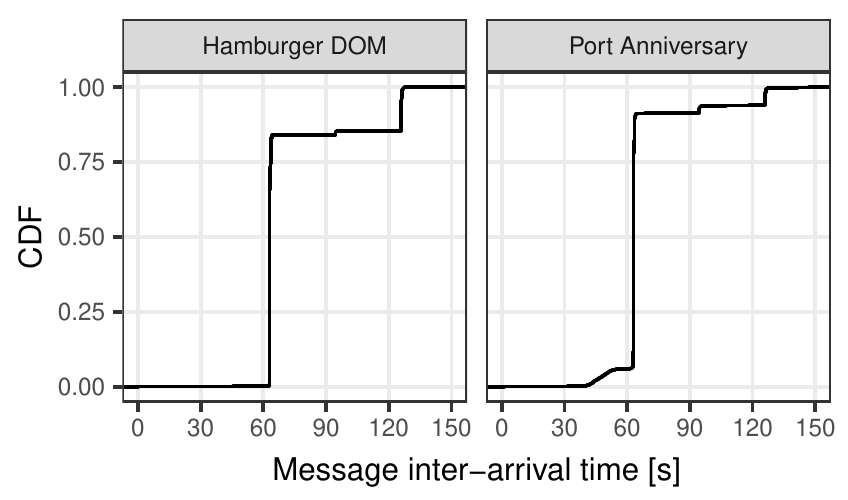}
	\caption{Comparison of message inter-arrival times for environmental sensors at the Hamburger DOM and Port Anniversary MONICA pilot event.}
	\label{fig:sensor_diff}
\end{figure}

\begin{figure}[t]
	\includegraphics[width=\columnwidth]{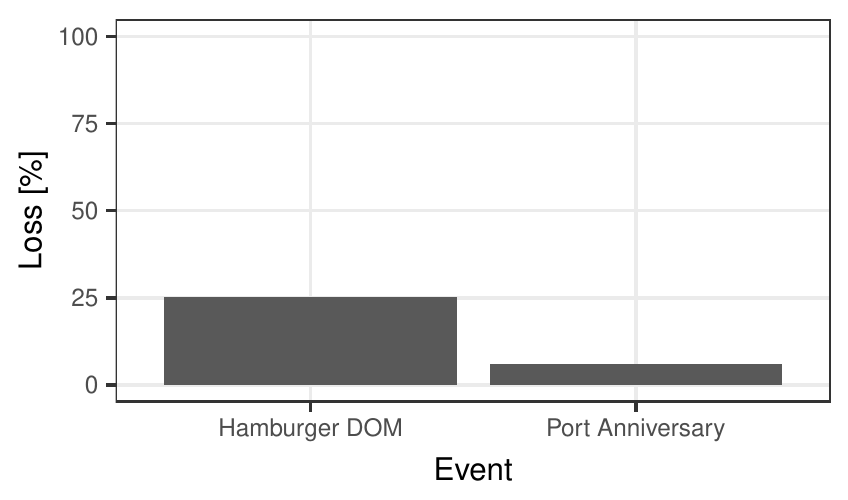}
	\caption{Comparison of message loss for environmental sensors at the Hamburger DOM and Port Anniversary MONICA pilot event.}
	\label{fig:sensor_loss}
\end{figure}

Next, we look at the measurement results for the environmental sensors, that were deployed at the Hamburg DOM funfair and the Port Anniversary.
The major difference between these deployments are the urban surroundings, i.e., the DOM festival site is in the dense city center with more and higher buildings nearby, while the port area is wider and less obscured by buildings.

This has direct effect on the number of LoRaWAN gateways reachable by a sensor during the events.
In \autoref{fig:sensor_dist} we compare the gateway-to-device distance and the relative amount of messages received by a certain gateway.
The results show that there are fewer gateways reachable around the DOM area and that these exhibit a significantly lower reception rate compare to the gateways in the port area.
However, this does not have a severe effect on the message inter-arrival times shown in \autoref{fig:sensor_diff}.
Over 80\,\% of the messages are received within the 1st interval at the DOM, compared to $\approx$ 90\,\% at the Port festival.
Which is also reflected by the summarised packet loss shown in \autoref{fig:sensor_loss}.
The results show that for both events (close to) 100\,\% of the sensor messages are received within 120\,s, considering the use-case this is acceptable.
Notably, we saw quite stable reception rate of 75\,\% for a gateway 14\,km away from the Port Anniversary deployment site.

In summary, our results show that a LoRaWAN infrastructure is able to support IoT applications with different requirements.
Specifically, our field deployments indicate high message receptions rates to even suffice near-realtime requirements of tracking/location services -- given low mobility. 

\section{Conclusion and Outlook}
\label{sec:outlook}

In this paper we presented the implementation, deployment, and evaluation of 2 Smart City applications based on LoRaWAN, the Open Source IoT operating system RIOT, and the IoT platform of the EU project MONICA.
Our results show that LoRaWAN is a suitable communication technology for many typical IoT scenarios - moreover, even when using a public infrastructure like TheThingsNetwork.
Further, LoRaWAN is able to handle near-realtime demands e.g. for accurate GPS tracking; given small, optimised data messages and slow velocities.
Even in dense city environments LoRaWAN achieves good radio coverage within a radius of 5\,km, and even beyond -- our results show repetitive, successful data transmission of up to 14\,km.

In our ongoing and future work we focus on long term IoT deployments and measurement studies in the city of Hamburg.
We aim to monitor the LoRaWAN coverage and evaluate the network performance considering e.g. varying LoRa configurations and payload sizes.
For instance, we currently run a test deployment of environmental sensors on the roof of buses of the public transport system.

%%% acks
\section*{Acknowledgment}

This work is co-funded by the European Union (EU) within the MONICA project under grant agreement number 732350. 
The MONICA project is part of the EU Framework Programme for Research and Innovation \emph{Horizon 2020} and the IoT European Large-Scale Pilots Programme.

\balance

%%% bibliography
\bibliographystyle{IEEEtran}
\bibliography{layer2,local,iot,own,internet,rfcs,ids}

\end{document}